\documentclass[preprint,12pt]{elsarticle}

\usepackage{amssymb,lineno,hyperref}

\journal{}
\begin{document}

\begin{frontmatter}



\title{Effects of the size and concentration of depleting agents on the stabilization of the double-helix structure and DNA condensation: a single molecule force spectroscopy study.}


\author{R. M. de Oliveira}
\address{Departamento de F\'isica, Universidade Federal de
	Vi\c{c}osa. Vi\c{c}osa, Minas Gerais, Brazil.}

\author{M. S. Rocha}
\address{Departamento de F\'isica, Universidade Federal de
Vi\c{c}osa. Vi\c{c}osa, Minas Gerais, Brazil.}
\ead{marcios.rocha@ufv.br}

\begin{abstract}

We perform a single molecule force spectroscopy study to characterize the role of the size (molecular weight) and concentration of depleting agents on DNA condensation and on the stabilization of the double-helix structure, showing that important features such as the threshold concentration for DNA condensation, the force in which the melting plateau occurs and its average length strongly depend on the depletant size chosen. Such results are potentially important to understand how the presence of surrounding macromolecules influences DNA stabilization inside living cells and therefore advance in the understanding of the crowded cell environment on DNA-related functions. 

\end{abstract}

\begin{keyword}
Polyethylene-glycol (PEG) \sep DNA \sep single molecule force spectroscopy \sep optical tweezers \sep depletion interactions

\end{keyword}

\end{frontmatter}

\section{INTRODUCTION}

Macromolecular crowding is an important phenomenon related to the presence of a relatively high concentration of macromolecules in solution, which in general drastically modify its properties \cite{EllisCrowding}. In fact, the presence of such macromolecules at high concentrations reduces the available volume inside solutions, interfering on the solvent activity and the thermodynamics of all other solute components. From a biological perspective, the intracellular medium is a crowded environment due to the presence of many types of proteins, sugars, nucleic acids and others. These molecules interfere in the relevant chemical reactions that occur inside cells, including the interactions involving nucleic acids, proteins and ligands \cite{Rocha2014, LimaBSAdoxo}. Therefore, to fully understand such interactions, they should be preferentially studied \textit{in vitro} using a crowded solution that mimics the intracellular medium.

In particular, when the crowding macromolecules are relatively large, depletion interactions can play a fundamental role in many colloidal and polymer solutions, modulating the general behavior of solute particles diluted in a given solvent containing these ``depletant agents''.  In general, such depletants present intermediate sizes between the solute and the solvent molecules, promoting relevant volume-exclusion effects and thus allowing the depletion phenomena to occur efficiently \cite{MaoDepletion, LekkerkerkerBook}. They can lead for example to particle aggregation and consequently to phase separation phenomena \cite{Asakura, MaoDepletion}. Common depletant agents used in various \textit{in vitro} studies are neutral polymers, proteins, micelles, osmolytes and others \cite{MaoDepletion, ButtBook}. 

Concerning specifically DNA solutions, it is well known that depletants such as neutral polymers \cite{LermanDepletion, VasilevyskayaPEG, RenkoDepletion, RamosJPC} and proteins \cite{Yoshikawa2010, RenkoDepletion, LimaBSA} can promote relevant depletion interactions between different DNA segments and, depending on the depletant concentration and ionic composition of the solution, lead to the aggregation of different DNA molecules (in concentrated DNA solutions) or, alternatively, to the collapse of a single DNA molecule by effective segment-segment attraction (in dilute DNA solutions), a phenomenon known as ``polymer-salt-induced'' (psi, $\psi$) condensation \cite{VasilevyskayaPEG, RenkoDepletion, FrischPsi}.

DNA $\psi$-condensation is a very important phenomenon from various different perspectives. From the biological point of view, the high concentration of depletants around collapsed DNA molecules in solution simulates two key features found inside cells: the crowded aspect of such environment as well the compacted DNA shape itself, which is somewhat similar to what occurs in prokaryotes \cite{Murphy, Murphy2, LimaBSAdoxo}. From the physicochemical point of view, on the other hand, such a system can be used for studying \textit{in vitro} interesting phenomena, \textit{e. g.}, the coil-globule transition that results in the condensation process itself \cite{Bloomfield, Bloomfield2} and furthermore, the effects of DNA condensation on its interactions with other compounds such as chemotherapeutic drugs \cite{Rocha2014, LimaBSAdoxo}. 

Although many aspects behind DNA $\psi$-condensation and depletion interactions (the driving force behind the phenomenon) are nowadays well understood and characterized, to the best of our knowledge a study concerning the role of depletion interactions on the stabilization of the secondary structure of the double-helix, performed at the single molecule level, is still lacking. Furthermore, the effects of the size of the depleting agents on the condensation process itself were not characterized at the single molecule level either, but only in a few bulk studies under specific conditions \cite{VasilevyskayaPEG, RamosJPC}. Here we fill these gaps by performing single molecule force spectroscopy with DNA molecules under various different crowded solutions, using optical tweezers (OT). We use the neutral polymer polyethylene-glycol (PEG) with different molecular weights as the depletant molecule, varying its concentration in a fixed solvent (a phosphate buffered saline (PBS) buffer). The experiments have shown that the depletant presents a strong size-dependent tendency to condense DNA and to stabilize the double-helix structure, being able to hinder the well-known force-induced melting plateau that occurs at $\sim$ 65 pN depending on its molecular weight and concentration. Such results are potentially important to understand how the presence of surrounding macromolecules influences DNA stabilization inside living cells and therefore advance in the understanding of the crowded cell environment on DNA-related functions.

\section{MATERIALS AND METHODS}

The samples prepared for single molecule OT assays consist of biotin-labeled $\lambda$-DNA (48,502 base-pairs, $\sim$ 16.5 $\mu$m contour length, New England Biolabs) attached by the ends to a streptavidin-coated polystyrene bead (3 $\mu$m diameter, Bangs Labs) and to a streptavidin-coated glass coverslip (used to construct the sample chamber). The samples are firstly prepared only with the DNA molecules in a Phosphate Buffered Saline (PBS) buffer composed of 4.375 mM of Na$_2$HPO$_4$, 1.25 mM of NaH$_2$PO$_4$ and 140 mM of NaCl (ionic strength $I$ = 154 mM). 

The experimental procedure is then performed as follows. Firstly, a particular DNA molecule is chosen and tested for integrity by performing force-extension measurements in the low-force entropic regime ($<$ 5 pN), measuring the persistence and contour lengths to verify if the values of such parameters are within the expected values \cite{RochaAPLPso, Alves2015}. Then, PEG is introduced in the sample chamber at a desired concentration, and we wait at least 20 minutes for equilibration. Finally, the same DNA molecule previously tested is stretched from an initial extension of 5 $\mu$m until reaching high stretching forces in the enthalpic (elastic) regime (some tens of pN), verifying how the presence of PEG in the sample modifies the expected force-extension curve of the biopolymer and, in particular, the melting plateau found at $\sim$ 65 pN for bare DNA molecules in our buffer.

\section{RESULTS AND DISCUSSION}

\begin{figure}
	\centering
	\includegraphics[width=10cm]{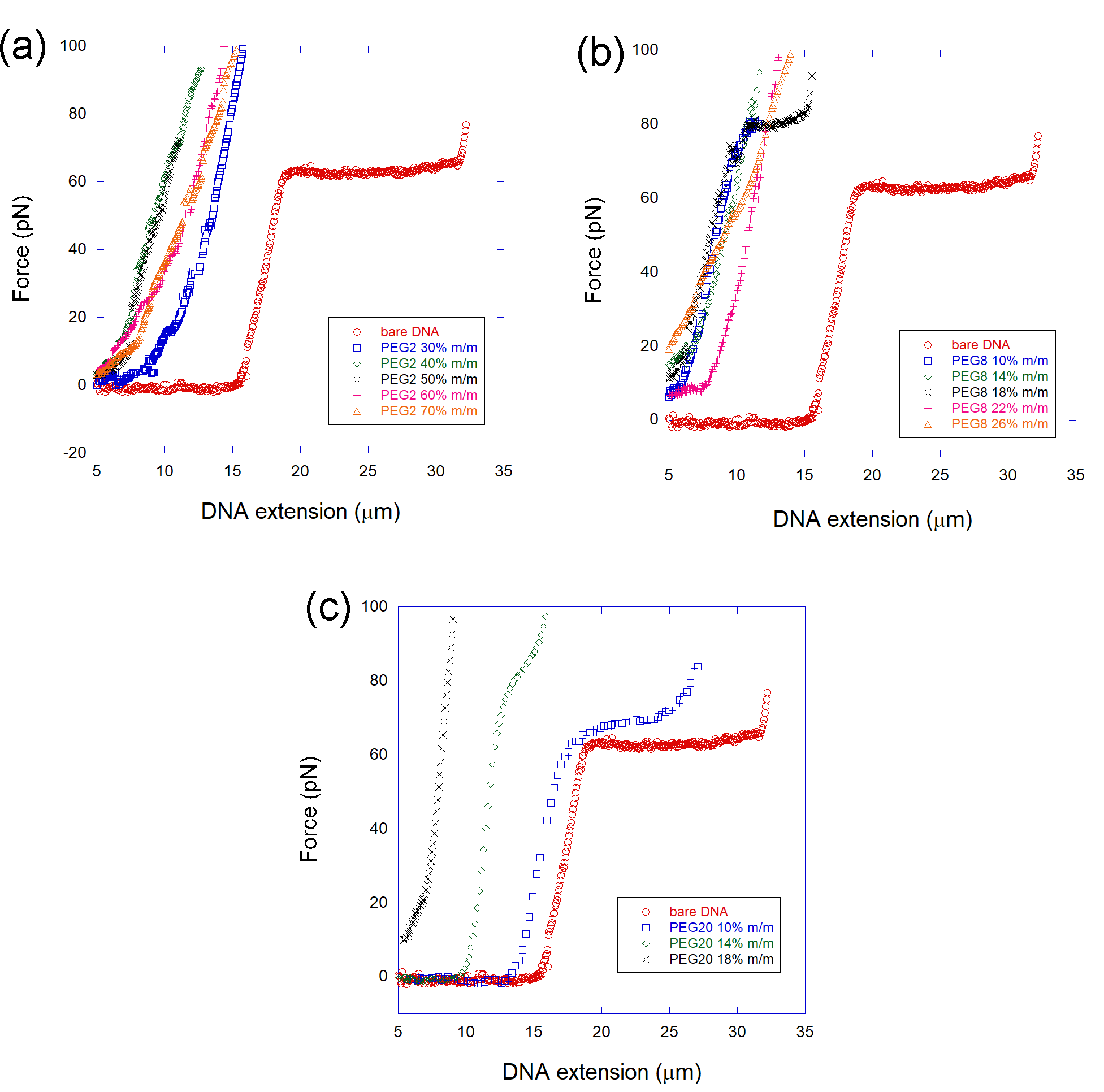}
	\caption{Representative force-extension curves (FECs) for various different PEG concentrations with molecular weights of $\sim$ 2,000 g/mol (PEG2k, panel \textit{a}), $\sim$ 8,000 g/mol (PEG8k, panel \textit{b}) and $\sim$ 20,000 g/mol (PEG20k, panel \textit{c}). The PEG concentrations are expressed here as mass fractions ($\%$ m/m) between the PEG mass and the total solution mass.}
	\label{FECs}
\end{figure}

In Fig. \ref{FECs} we show some representative force-extension curves (FECs) for various different PEG concentrations with molecular weights of $\sim$ 2,000 g/mol (PEG2k, panel \textit{a}), $\sim$ 8,000 g/mol (PEG8k, panel \textit{b}) and $\sim$ 20,000 g/mol (PEG20k, panel \textit{c}). The PEG concentrations are expressed here as mass fractions ($\%$ m/m) between the PEG mass and the total solution mass. Observe that for bare $\lambda$-DNA (without PEG in the sample) the FEC exhibits the expected behavior well described by the Worm-Like Chain (WLC) model, with the melting plateau at $\sim$ 65 pN, where the biopolymer have its contour length increased by about $\sim$ 1.7$\times$ during the melting transition (\textit{red circles} in all panels). 

For PEG2k and PEG8k the qualitative behavior is similar: the FECs readily lose the melting plateau, which becomes rare, appearing sporadically only in a few curves; and in those cases much smaller in ``length'' (the plateau horizontal extension, in micrometers) and much higher in ``height'' (the average force where the plateau occurs, in pN) than the results found for bare DNA ($\sim$ 12 $\mu$m length, $\sim$ 65 pN), as can be noted in Fig. \ref{FECs}\textit{a} and \textit{b}. Furthermore, observe that the maximum DNA extension reached for a given force decreases when PEG is present, confirming the condensation phenomenon. In the case of PEG2k (Fig. \ref{FECs}\textit{a}), the condensation typically occurs in the PEG concentration range of 40-70$\%$ m/m, much higher than the corresponding range found for PEG8k (Fig. \ref{FECs}\textit{b}), which is 10-26$\%$ m/m. Such result is in agreement with previous single molecule and bulk studies performed with various types of DNA depletants \cite{Yoshikawa2010, LimaBSA, CrisafuliPEG} and attest the important role of the PEG size on the DNA condensation process. In the case of PEG20k (Fig. \ref{FECs}\textit{c}), the behavior is qualitatively different: here the melting plateau is much more common and disappears gradually as the PEG concentration is increased in the sample. At 10$\%$ of PEG20k (\textit{blue squares} in Fig. \ref{FECs}\textit{c}), observe that the melting plateau already starts to disappear: the average force needed to promote DNA melting increases and the plateau length decreases, indicating that the melting process had become more difficult to occur. Such situation is intensified for higher PEG20k concentrations (see for example 14$\%$ m/m; \textit{green diamonds} in Fig. \ref{FECs}\textit{c}) and, finally, at 18$\%$ of PEG20k the plateau completely disappears, showing that the PEG-induced depletion interactions have hindered the force-induced melting transition. Observe that PEG20k also promoted the expected DNA compaction, with the measured extension for a given force decreasing with the PEG concentration in the sample, as can be clearly seen in Fig. \ref{FECs}\textit{c}.  

\begin{figure}
	\centering
	\includegraphics[width=10cm]{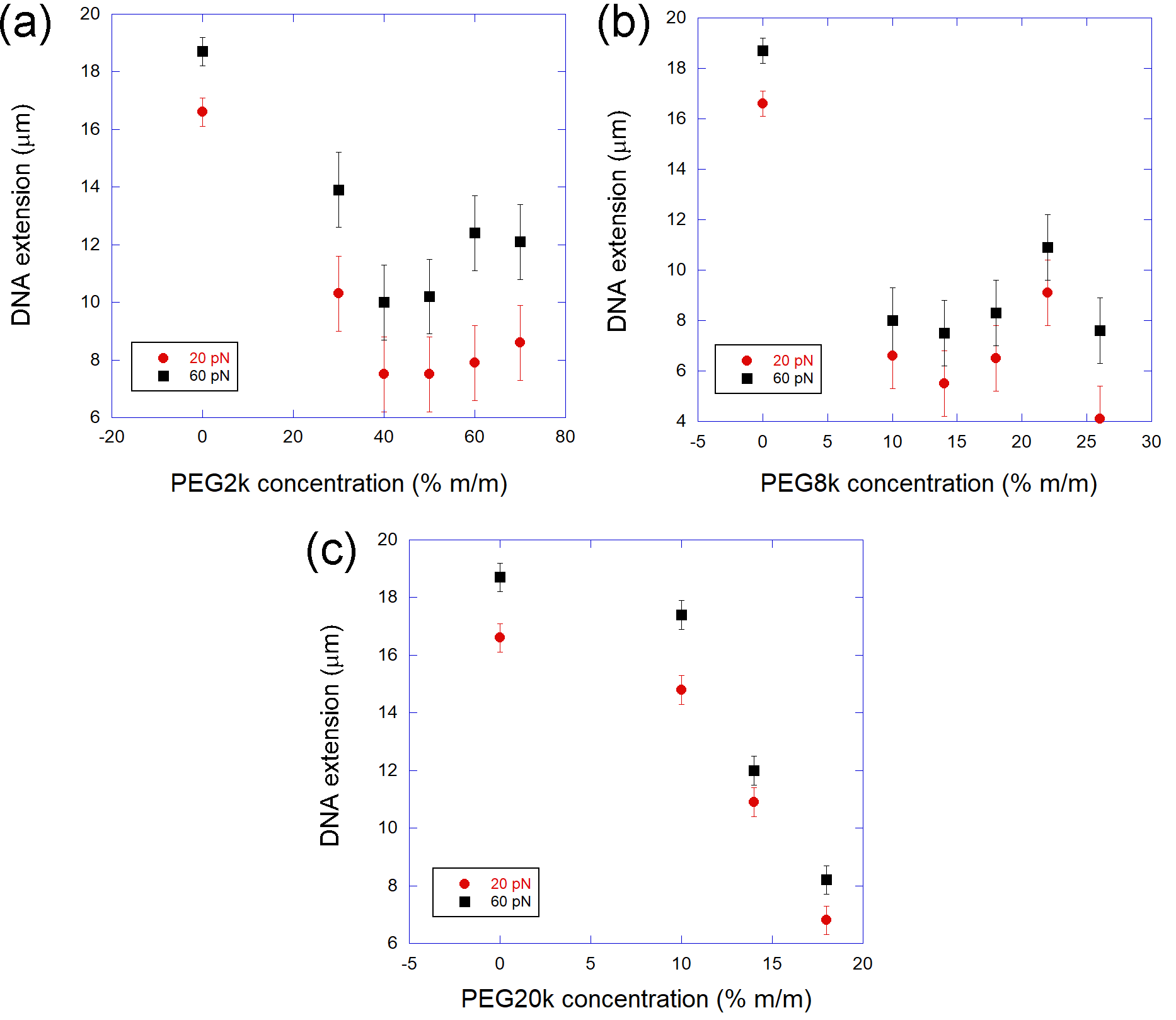}
	\caption{DNA extension measured at two characteristic forces (20 pN and 60 pN) as a function of PEG concentration for the three different PEGs used: PEG2k (panel \textit{a}), PEG8k (panel \textit{b}) and PEG20k (panel \textit{c}). These data explicitly show the very different qualitative behavior concerning the DNA condensation process induced by PEG: for the two PEGs with lower molecular weights (2k and 8k) the DNA extension decreases abruptly as a function of the PEG concentration, remaining practically constant for concentrations $>$ 40$\%$ m/m (PEG2k) and $>$ 10$\%$ m/m (PEG8k). For PEG20k, on the other hand, the decrease of the DNA extension as a function of the PEG concentration is much more smooth, occurring gradually.}
	\label{zC}
\end{figure}

In order to get a better overview on the DNA condensation process promoted by PEG under the different studied situations, in Fig. \ref{zC} we show the DNA extension measured at two characteristic forces (20 pN and 60 pN) as a function of the PEG concentration for the three different PEGs used: PEG2k (panel \textit{a}), PEG8k (panel \textit{b}) and PEG20k (panel \textit{c}). Such data was directly extracted from the measured FECs (whose some examples were already shown in Fig. \ref{FECs}). These data explicitly show the very different qualitative behavior already mentioned concerning the DNA condensation process by PEG: for the two PEGs with lower molecular weights (2k and 8k) the DNA extension decreases abruptly as a function of the PEG concentration, remaining practically constant for concentrations $>$ 40$\%$ m/m (PEG2k, Fig. \ref{zC}\textit{a}) and $>$ 10$\%$ m/m (PEG8k, Fig. \ref{zC}\textit{b}). For PEG20k, on the other hand, the decrease of the DNA extension as a function of the PEG concentration is much more smooth, occurring gradually (Fig. \ref{zC}\textit{c}). Unfortunately, it was not possible to increase the PEG20k concentration anymore in order to clearly view a saturation in the DNA extension decrease, due to the limited solubility of this high molecular weight PEG.

\begin{figure}
	\centering
	\includegraphics[width=10cm]{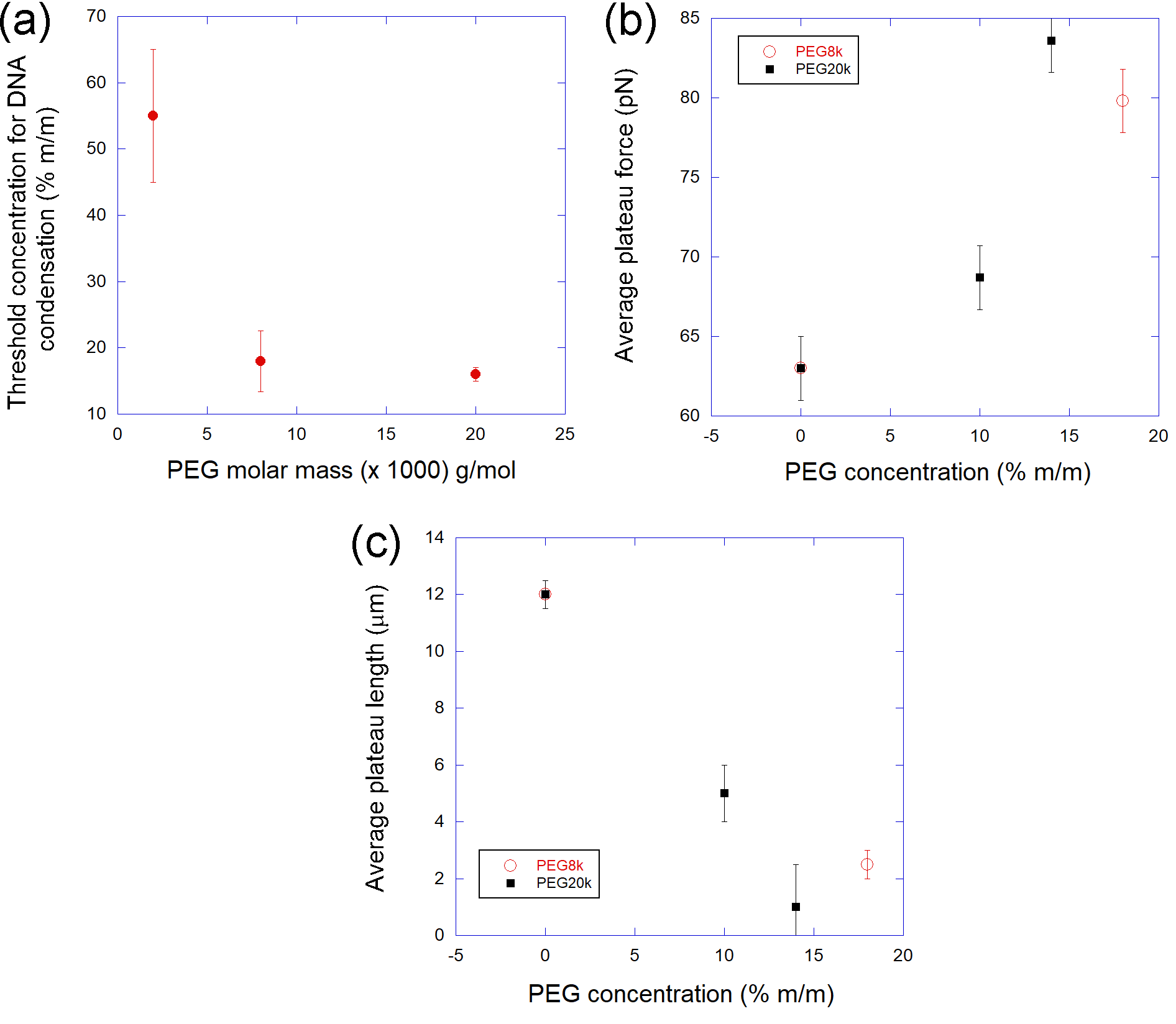}
	\caption{(\textit{a}) Average threshold concentration for DNA condensation (calculated as the average over the concentration range where DNA condenses) as a function of the PEG molecular weight. (\textit{b}) Average force in which the melting plateau occurs as a function of the PEG concentration. (\textit{c}) Average length of the melting plateau, in micrometers, as a function of the PEG concentration.}
	\label{AvgCurves}
\end{figure}

Finally, in Fig. \ref{AvgCurves} we show the average results found for some key quantities concerning the characterization of the PEG-induced DNA condensation process and the modification of the melting plateau already mentioned. Panel \textit{a} complements the analysis performed in the former figure, showing the average threshold concentration for DNA condensation (calculated as the average over the concentration range where DNA condenses) as a function of the PEG molecular weight. Observe that the higher the molecular weight, the lower the concentration one needs to condense DNA in solution, a result that corroborates with previous works concerning depletion interactions and DNA $\psi$-condensation \cite{VasilevyskayaPEG}. Panel \textit{b} shows the average force in which the melting plateau occurs as a function of the PEG concentration. We were able to obtain some data for PEG20k and PEG8k in this case. Observe that, independent on the molecular weight, the melting plateau tends to occur at higher forces when one increases the PEG concentration in the sample, indicating that the depletion interactions promoted by the neutral polymer in solution stabilizes the DNA double-helix structure, hindering the denaturation of the biopolymer. Such conclusion is confirmed by the data show in panel \textit{c}, where we plot the average length of the melting plateau, in micrometers, as a function of the PEG concentration. Observe that in this case such length rapidly decreases when the PEG concentration is increased, which shows that the denatured DNA portion strongly decreases for higher PEG concentrations, therefore confirming the stabilization of the double-helix structure against force-induced melting. Such a result also corroborates with studies performed using temperature-induced melting assays, which is the straightforward bulk method for characterizing the melting of double-stranded nucleic acids and have pointed that PEGs with molecular weights larger than 1,000 g/mol stabilizes DNA and RNA duplexes against temperature-induced melting \cite{NakanoPEG, PramanikPEG, KarimatakPEG, GuPEG}.

In summary, the results presented here allows us to draw some important new conclusions concerning the DNA $\psi$-condensation process and the stabilization of the double-helix structure against force-induced melting due to the depletion interactions promoted by the presence of PEG in solution. Key quantities such as the average threshold concentration for DNA condensation to occur, the average force in which the melting plateau occurs and the average length of this plateau strongly depend of the PEG size (molecular weight) chosen: higher molecular weight PEGs (20k g/mol) promote a more gradual condensation and melting stabilization, while lower molecular weight PEGs (2k and 8k g/mol) tend to interact more directly with DNA via depletion interactions, rapidly condensing the biopolymer and stabilizing the double-helix structure against force-induced melting, although higher PEG concentrations are needed in this case. This fact occurs because lower size PEGs promote weaker volume-exclusion effects and, although such a feature decreases the ability to condense DNA (requiring higher concentrations), it allows a higher number of molecules to stay close to the double-helix in solution, stabilizing this structure against melting.

\section{CONCLUSIONS}

We performed single molecule force spectroscopy with optical tweezers to study the effects of the size (molecular weight) and concentration of depleting agents on the DNA $\psi$-condensation and on the stabilization of the double-helix structure under force-induced melting. To perform such study we used the classic neutral polymer polyethylene-glycol (PEG) as the depleting agent and $\lambda$-DNA in a simple phosphate buffered saline solution with a fixed ionic strength of physiological relevance. The results achieved have shown that the intrinsic details related to the DNA condensation process (average threshold concentration for DNA condensation) and the stabilization of the double-helix structure (average force in which the melting plateau occurs, average length of this plateau) strongly depend of the PEG size chosen. In particular, higher molecular weight PEGs (20k g/mol) promote a more gradual condensation and melting stabilization, while lower molecular weight PEGs (2k and 8k g/mol) tend to interact more directly with DNA via depletion interactions, rapidly condensing DNA and stabilizing the double-helix structure against force-induced melting, although higher PEG concentrations are needed in this case.

\section{Acknowledgements}
This research was funded by Conselho Nacional de Desenvolvimento Cient\'ifico e Tecnol\'ogico (CNPq); Funda\c{c}\~ao de Amparo \`a Pesquisa do Estado de Minas Gerais (FAPEMIG); and Coordena\c{c}\~ao de Aperfei\c{c}oamento	de Pessoal de N\'ivel Superior (CAPES) - Finance Code 001.

\bibliography{PEG2023_bibtex}

\begin{thebibliography}{10}
\expandafter\ifx\csname url\endcsname\relax
  \def\url#1{\texttt{#1}}\fi
\expandafter\ifx\csname urlprefix\endcsname\relax\def\urlprefix{URL }\fi
\expandafter\ifx\csname href\endcsname\relax
  \def\href#1#2{#2} \def\path#1{#1}\fi

\bibitem{EllisCrowding}
R.~J. Ellis, Macromolecular crowding: obvious but underappreciated., Trends
  Biochem. Sci. 26~(10) (2001) 597--604.

\bibitem{Rocha2014}
M.~S. Rocha, A.~G. Cavalcante, R.~Silva, E.~B. Ramos, On the effects of
  intercalators in dna condensation: a force spectroscopy and gel
  electrophoresis study., J. Phys. Chem. B 118~(18) (2014) 4832--4839.

\bibitem{LimaBSAdoxo}
C.~H.~M. lima, H.~M.~C. de~Paula, L.~H.~M. da~Silva, M.~S. Rocha, Doxorubicin
  hinders dna condensation promoted by the protein bovine serum albumin (bsa).,
  Biopolymers 107~(12) (2017) e23071.

\bibitem{MaoDepletion}
Y.~Mao, M.~E. Cates, H.~N.~W. Lekkerkerker, Depletion force in colloidal
  systems, Physica A 222~(1-4) (1995) 10--24.

\bibitem{LekkerkerkerBook}
H.~N.~W. Lekkerkerker, R.~Tuinier, Colloids and the Depletion Interaction, 1st
  Edition, Springer: Heidelberg., 2011.

\bibitem{Asakura}
S.~Asakura, F.~Oosawa, On interaction between two bodies immersed in a solution
  of macromolecules., J. Chem. Phys. 22 (1954) 1255--1256.

\bibitem{ButtBook}
H.-J. Butt, K.~Graf, M.~Kappl, Physics and chemistry of interfaces, 2nd
  Edition, Weinheim: Wiley-VCH-Verl., 2006.

\bibitem{LermanDepletion}
L.~S. Lerman, A transition to a compact form of dna in polymer solutions.,
  Proc. Natl. Acad. Sci. USA 68~(8) (1971) 1886--1890.

\bibitem{VasilevyskayaPEG}
V.~V. Vasilevskaya, A.~R. Khokhlov, Y.~Matsuzawa, K.~Yoshikawa, Collapse of
  single dna molecule in poly(ethylene glycol) solutions., J. Chem. Phys.
  102~(16) (1995) 6595--6602.

\bibitem{RenkoDepletion}
R.~de~Vries, Dna compaction by nonbinding macromolecules, Polymer Sci. C 54~(1)
  (2012) 30--35.

\bibitem{RamosJPC}
E.~B. Ramos, R.~de~Vries, J.~R. Neto, Dna psi-condensation and reentrant
  decondensation: Effect of the peg degree of polymerization., J. Phys. Chem. B
  109~(49) (2005) 23661--5.

\bibitem{Yoshikawa2010}
K.~Yoshikawa, S.~Hirota, N.~Makita, Y.~Yoshikawa, Compaction of dna induced by
  like-charge protein: Opposite salt-effect against the polymer-salt-induced
  condensation with neutral polymer., Phys. Chem. Lett. 1 (2010) 1763--1766.

\bibitem{LimaBSA}
C.~H.~M. lima, H.~M.~C. de~Paula, L.~H.~M. da~Silva, M.~S. Rocha, Unfolding dna
  condensates produced by dna-like charged depletants: A force spectroscopy
  study., J. Chem. Phys. 146 (2017) 054901.

\bibitem{FrischPsi}
H.~L. Frisch, S.~Fesciyan, Dna phase transitions: The $\psi$ transition of single
  coils., J. Pol. Sci. Pol. Lett. Ed. 17~(5) (1979) 309--315.

\bibitem{Murphy}
L.~D. Murphy, S.~B. Zimmerman, Macromolecular crowding effects on the
  interaction of dna with escherichia coli dna-binding proteins: a model for
  bacterial nucleoid stabilization., Biochim. Biophys. Acta. 1219~(2) (1994)
  277--284.

\bibitem{Murphy2}
L.~D. Murphy, S.~B. Zimmerman, Condensation and cohesion of $\lambda$ dna in
  cell extracts and other media: Implications for the structure and function of
  dna in prokaryotes., Biophys. Chem. 57~(2) (1995) 71--92.

\bibitem{Bloomfield}
V.~A. Bloomfield, Dna condensation by multivalent cations, Biopolymers 44~(3)
  (1997) 269--282.

\bibitem{Bloomfield2}
V.~A. Bloomfield, Condensation of dna by multivalent cations: considerations on
  mechanism, Biopolymers 31 (1991) 1471--1481.

\bibitem{RochaAPLPso}
M.~S. Rocha, A.~D. Lucio, S.~S. Alexandre, R.~W. Nunes, O.~N. Mesquita,
  Dna-psoralen: Single-molecule experiments and first principles calculations,
  Appl. Phys. Lett. 95 (2009) 253703.

\bibitem{Alves2015}
P.~S. Alves, O.~N. Mesquita, M.~S. Rocha, Controlling cooperativity in
  $\beta$-cyclodextrin-dna binding reactions., J. Phys. Chem. Lett. 6 (2015)
  3549--3554.

\bibitem{CrisafuliPEG}
F.~A.~P. Crisafuli, L.~H.~M. da~Silva, G.~D.~M. Ferreira, E.~B. Ramos, M.~S.
  Rocha, Depletion interactions and modulation of dna-intercalators binding:
  Opposite behavior of the ``neutral'' polymer poly(ethylene-glycol).,
  Biopolymers 105~(4) (2015) 227--233.

\bibitem{NakanoPEG}
S.~ichi Nakano, H.~Karimata, T.~Ohmichi, J.~Kawakami, N.~Sugimoto, The effect
  of molecular crowding with nucleotide length and cosolute structure on dna
  duplex stability., J. Am. Chem. Soc. 126~(44) (2004) 14330--14331.

\bibitem{PramanikPEG}
S.~Pramanik, S.~Nagatoishi, S.~Saxena, J.~Bhattacharyya, N.~Sugimoto,
  Conformational flexibility influences degree of hydration of nucleic acid
  hybrids., J. Phys. Chem. B 115~(47) (2011) 13862--13872.

\bibitem{KarimatakPEG}
H.~Karimata, S.~ichi Nakano, N.~Sugimoto, Effects of polyethylene glycol on dna
  duplex stability at different nacl concentrations., Bull. Chem. Soc. Jpn.
  80~(10) (2007) 1987--1994.

\bibitem{GuPEG}
X.~Gu, M.-T. Nguyen, A.~Overacre, S.~Seaton, S.~J. Schroeder, Effects of salt,
  polyethylene glycol, and locked nucleic acids on the thermodynamic
  stabilities of consecutive terminal adenosine mismatches in rna duplexes., J.
  Phys. Chem. B 117~(13) (2013) 3531--3540.

\end{thebibliography}

\end{document}